\documentstyle[12pt]{article}
\begin{document}

\begin{titlepage}
\begin{center}

\bigskip

\bigskip

\vspace{3\baselineskip}

{\Large \bf  Scale Invariant Kaluza-Klein Theory\\

\bigskip

  and the Fate of the Gravitational Collapse}

\bigskip

\bigskip

\bigskip

\bigskip

{\bf  Israel Quiros}\\
\smallskip

{ \small \it  
Physics Department. Las Villas Central University.\\
Santa Clara . Villa Clara. Cuba}

\bigskip

{\tt  israel@uclv.edu.cu} 

\bigskip

\bigskip

\bigskip

\bigskip

\vspace*{.5cm}

{\bf Abstract}\\
\end{center}
\noindent
Pushing forward the similitudes between the gravitational collapse and the expansion of the universe (in the reversed sense of time), it should be expected that, during the collapse, eventually, a spacetime domain would be reached where attained energy scales are very high. In consequence some of the compactified extra dimensions may be decompactified and some presently broken symmetries may be restored. A more fundamental theory (of which Einstein's theory is a symmetry broken phase) is then expected to take account of further description of the collapse. I propose a simple (classical) model for the description of the late stages of the gravitational collapse: A non-Riemannian, scale-invariant version of 5-dimensional Kaluza-Klein theory in which the standard Riemann structure of the higher-dimensional manifold is replaced by a Weyl-integrable one. A class of solutions, that generalize the "soliton" one by Gross and Perry and Davidson and Owen, is found. This class contains both naked singularities and wormhole solutions. On physical grounds it is argued that a wormhole is the most reasonable destiny of the gravitational collapse. 


\bigskip

\bigskip

\end{titlepage}

\section{Introduction}

The issue of the spacetime singularities has been intensively studied in the past. It has been shown that spacetime singularities inevitably occur if the energy content holds some reasonable energy conditions\cite{hawking}. However, since the presence of these undesirable geometrical objects means breakdown of physical predictability, it is usually argued that the singularities are artifacts of general relativity -being a classical construction- due to the lack of quantum considerations. The fact is that, almost 90 years after the first papers on general relativity appeared and more than a century after the birth of quantum theory, we have not yet a manageable quantum theory of spacetime. In recent years much work has been aitmed at both classical and semiclassical description of non-singular behaviour\cite{singularity,novello}. 

A picture without singularity is physically justified if one adheres to the ideology of the unification scheme of the fundamental interactions. In effect, according to this scheme, during the early stages of the evolution the universe was higher-dimensional and the physical laws were maximally symmetric. Subsequently, as a result of decreasing of the universe's temperature, some of the extra dimensions were compactified and some symmetries of the laws of physics were expontaneously (or dynamically) broken. This evolution pattern led to the present four-dimensional world (according to the brane models\cite{arkani-hamed} it could be well higher-dimensional) with some broken symmetries. Regarding dimensionality, there are convincing arguments showing that there was a stage in the evolution of the universe when it was five-dimensional\cite{horava-witten}. Respecting symmetry breaking, it is well known that break down of conformal symmetry is usually linked with mass generating mechanisms in particle physics. Therefore, the fact that, at present, some particles have non-null masses is the best evidence we have that the laws of physics were conformally invariant in the past. Thinking along these lines one is led to conclude that Einstein's gravity is a conformal-symmetry-broken phase of a more fundamental (conformal) theory. M-theory\cite{duff} (of which 11-dimensional supergravity is the low-energy limit) and (10-dimensional) string theory\cite{witten} seem to be the best candidates. This more fundamental theory could be able to remove the spacetime singularities from the description of the physical world.

If we push forward the similitudes between the expansion of the universe (in the reversed sense of time) and the gravitational collapse\cite{hawking}, then one is led to conclude that, in the late stages of the collapse, a spacetime domain could be eventually reached where some presently broken symmetries are restored and some of the compact dimensions could decompactify. In such a domain four-dimensional Einstein's gravity could be replaced by a higher-dimensional conformal theory which could take account for further description of the collapse. In the present paper I study a very simple classical model in which the geometrical structure of the five-dimensional Kaluza-Klein (KK) theory\cite{kk} is modified by enlarging the symmetry group of the 5-dimensional Einstein-Hilbert action $S_5=\int_{M_5^R}\frac{d^5x\sqrt{|g|}}{16\pi G_5}\hat R$, where $g$ is the determinant of the 5-dimensional metric $g_{AB}$, $\hat R$ is the 5-dimensional Ricci scalar, $G_5$ stands for the 5-dimensional Newton's constant and $M_5^R$ is a 5-dimensional Riemann manifold, to include scale invariance. I assume, besides that the underlying geometric structure of the theory is a Weyl-integrable one\cite{novello,weyl}. The role of structural phase transitions of the underlying geometry\cite{novello} is usually diminished: If Einstein's gravity is a conformal-symmetry broken phase of a more fundamental theory then, Riemann geometry should be a conformal-symmetry broken phase of a more fundamental geometry that, presumably, dominated the early stages of the cosmological evolution. Weyl geometry seems to be a good alternative since, as pointed out in the bibliography\cite{novello,dzhunushaliev}, quantum behaviour could be obtained from a classical formalism based on Weyl spaces. The possibility that higher-dimensional KK-gravity may account for vanishing of spacetime singularities was advanced in Ref.\cite{sato}. In that paper Tomimatsu studied 5-dimensional KK-theory in the "neutral" matter gauge (no vector gauge fields present) and showed that, what is usually understood in four dimensions as a spherically-symmetric Schwarzschild black hole, could be interpreted as a "window" to the extra dimension. It is precisely the kind of picture one expects is to take place, if the timelike singularity hidden inside the black hole even horizon is to be removed from the physical description of the world.

The model I want to present will be explained in the next Section. In Section {\bf 3} I will outline the conformal technique that will be used in Sec. {\bf 4} to generate solutions to this model, from the known "soliton" class of solution of Ref.\cite{wesson,gross-perry}. The results will be disscused in Sec. {\bf 5} and brief conclusions given in the last ({\bf 6}th) Section. 

\section{The Model}

The modified, scale-invariant 5-dimensional action being proposed is the following,

\begin{equation}
\label{action}
S_5^W =\int_{M_5^W}\frac{d^5x\sqrt{|g|}}{16\pi G_5}e^{-\frac{3}{2}\omega}R, 
\end{equation}
where $M_5^W$ is a 5-dimensional Weyl-integrable manifold given by the pair $(g_{AB},\omega)$, $\omega$ being a Weyl scalar function and the uppercase latin indexes $A,B=\overline{0,3},5$. The requirement that the covariant derivative of the metric tensor is non-vanishing: $\nabla_C g_{AB}=\omega_{,C}g_{AB}$ should be fulfilled. $R=g^{MN}R_{MN}$, $R_{AB}=\Gamma_{AB,N}^N-\Gamma_{NB,A}^N\+\cdots$, with the manifold affine connection defined as $\Gamma_{AB}^C=\{^{\;C}_{AB}\}-\frac{1}{2}(\omega_{,A}\delta_B^C+\omega_{,B}\delta_A^C-g_{AB}\omega^{,A})$, and $\{^{\;C}_{AB}\}=\frac{1}{2}g^{CN}(g_{AN,B}+g_{BN,A}-g_{AB,N})$ are the standard 5-dimensional Christoffel symbols of the metric $g_{AB}$. This action is invariant under the Weyl rescalings

\begin{equation}
\label{rescalings}
g_{AB}\rightarrow\Omega^2 g_{AB},\;\;\omega\rightarrow\omega+\ln\Omega^2,
\end{equation}
where $\Omega^2$ is a smooth function on the manifold. In terms of Riemannian magnitudes and operators (in what follows these are distinguished with a hat) that are defined in respect to the Christoffel symbols of the metric: $\hat R_{AB}=\{^{\;N}_{AB}\}_{,N}-\{^{\;N}_{NB}\}_{,A}+\cdots$, $\hat\Box\xi=g^{MN}\hat\nabla_M\hat\nabla_N\xi=g^{MN}(\xi_{,MN}-\{^{\;\;S}_{MN}\}\xi_{,S})$, etc., and if one rescales the Weyl scalar function $\phi=e^{-\frac{3}{2}\omega}$, the action (\ref{action}) can be written in a Brans-Dicke form with coupling parameter $-\frac{4}{3}$:

\begin{equation}
\label{bdaction}
S_5^W =\int_{M_5^W}\frac{d^5x\sqrt{|g|}}{16\pi G_5}\{\phi\hat R+\frac{4}{3}\phi^{-1}(\hat\nabla\phi)^2\},
\end{equation}
where $(\hat\nabla\phi)^2\equiv g^{MN}\hat\nabla_M\phi\hat\nabla_N\phi=g^{MN}\phi_{,M}\phi_{,N}$. The field equations derivable from (\ref{action}) are the following:

\begin{eqnarray}
\label{fieldeq}
\hat R_{AB}&=&\phi^{-1}\hat\nabla_A\hat\nabla_B\phi-\frac{4}{3}\frac{\phi_{,A}}{\phi}\frac{\phi_{,B}}{\phi},\nonumber\\
\hat\Box\phi&=&0.
\end{eqnarray}

The theory outlined here is, like the standard KK-theory, or the induced matter approach of Wesson\cite{wesson}, pure geometry in the higher dimension. Recall that $\phi$ (or, instead $\omega$) is not a matter field since it enters in the definition of the manifold affine connection. This approach is distinguished by invariance under the Weyl rescalings (\ref{rescalings}) and non-Riemannian (Weyl-integrable) geometric structure of manifolds that solve (\ref{fieldeq}).

\section{The Conformal Technique}

Now I apply the conformal technique for generating solutions to the above theory. The first step is to perform the following conformal transformation in (\ref{bdaction}): $g_{AB}\rightarrow\phi^{-\frac{2}{3}}g_{AB}$, to obtain: $S_5=\int_{M_5^R}\frac{d^5x\sqrt{|g|}}{16\pi G_5}\hat R$, where we made evident the fact that, under the above conformal transformation $\Gamma_{AB}^C\rightarrow\{^{\;C}_{AB}\}$, and $\hat\nabla_C g_{AB}=-\frac{2}{3}\frac{\phi_{,C}}{\phi}g_{AB}\rightarrow\hat\nabla_C g_{AB}=0$. I. e., one recovers Riemannian structure on the manifold ($M_5^W\rightarrow M_5^R$). In consequence we obtain the 5-dimensional Einstein-Hilbert action for pure gravity (vacuum) so, the equations (\ref{fieldeq}) are mapped into the following equations,

\begin{eqnarray}
\label{fieldeq1}
\hat R_{AB}&=&0,\nonumber\\
\frac{\hat\Box\phi}{\phi}&=&\frac{(\hat\nabla\phi)^2}{\phi^2}.
\end{eqnarray} 

The next step is to find the pair ($g_{AB},\phi$) that solves (\ref{fieldeq1}). Therefore, the original scale-invariant metric on $M_5^W$, that is solution of (\ref{fieldeq}), can be found by applying the inverse conformal transformation, i. e., the 5-dimensional Weylian line-element that solves (\ref{fieldeq}) is the following:

\begin{equation}
\label{line}
ds_5^2=\phi^\frac{2}{3} g_{MN}dx^M dx^N,
\end{equation}
where the $g_{AB}$ are solutions of the first equation in (\ref{fieldeq1}), meanwhile $\phi$ is solution of the second one.

\section{The Soliton-like Class of Solutions}

I want to apply this theory to discuss the spherically symmetric gravitational collapse. For this reason I will be interested in the spherically symmetric (Schwarzschild) class of solutions to (\ref{fieldeq1}), also acknowledged as soliton solutions\cite{gross-perry}. The solitonic line-element can be written in the following form\cite{wesson},

\begin{equation}
\label{soliton}
ds_5^2=-A^adt^2+A^{-(a+b)}dr^2+A^{1-(a+b)}r^2d\Omega^2+A^bdy^2,
\end{equation}
where $d\Omega^2=d\theta^2+\sin^2\theta d\varphi$, $y$ stands for the extra coordinate and $a$ and $b$ are dimensionless constants related by the consistency relation $a^2+ab+b^2=1$. In case the cilindricity condition holds\cite{wesson,gross-perry}, $A=1-2M/r$. In the particular case when $a=1$, $b=0$, we recover 4-dimensional Schwarzschild solution with an extra flat dimension. I will consider a more general situation in which cilindricity is relaxed. Besides, like in Wesson's approach\cite{wesson}, no apriori assumption on the topology of the extra dimension is made. Then in the line-element (\ref{soliton}) I assume $A=A(r,y)$ and, besides, $\phi=\phi(r,y)$. I will be interested in solutions to (\ref{fieldeq1}) that are separable in $r$ and $y$, i. e., $A(r,y)=\alpha(r)\beta(y)$ and $\phi(r,y)=\xi(r)\zeta(y)$. In consequence, the field equations (\ref{fieldeq1}) are split into the following set of equations,

\begin{equation}
\label{split1}
\frac{\alpha''}{\alpha}+\frac{2}{r}\frac{\alpha'}{\alpha}=0,\;\;\;
\frac{\xi''}{\xi}+(\frac{2}{r}+\frac{\alpha'}{\alpha})\frac{\xi'}{\xi}-(\frac{\xi'}{\xi})^2=0,
\end{equation}
and,

\begin{equation}
\label{split2}
\frac{\ddot\beta}{\beta}-(a+2b)(\frac{\dot\beta}{\beta})^2=0,\;\;\;
\frac{\ddot\zeta}{\zeta}+[1-(a+b)]\frac{\dot\beta}{\beta}\frac{\dot\zeta}{\zeta}-(\frac{\dot\zeta}{\zeta})^2=0,
\end{equation}
where the comma denotes derivative in respect to the $r$-coordinate while the dot stands for derivative in respect to the $y$-coordinate. Equations (\ref{split1}) can be readily integrated to yield: $\alpha=1-2M/r$ and $\xi=(1-2M/r)^{k/2M}$, where $M$ and $k$ are integration constants ($M$ can be interpreted as the mass located at the center of the 3-space). Integration of the second pair of equations (\ref{split2}) yields, for $a\neq 1$, $b\neq 0$, $\beta=h(l-y)^{-\frac{1}{a+2b-1}}$ and $\zeta=(l-y)^m$, where $h$, $l$, and $m$ are integration constants. In the case when $a=1$, $b=0$, (properly the Schwarzschild case) one has, $\frac{\dot\beta}{\beta}=const.\rightarrow\beta=e^{ly}$ and $\frac{\dot\zeta}{\zeta}=const.\rightarrow\zeta=e^{hy}$ respectively. If one now applies the conformal technique to go back to the original Weyl-integrable manifold (see Sec. {\bf 3}), then one is faced with the following expressions for the Weyl-integrable line-element that solves (\ref{fieldeq}): For the case when $a\neq 1$ and $b\neq 0$,

\begin{equation}
\label{sol1}
ds_5^2=-\frac{h^a(l-y)^{\bar m}}{(1-\frac{2M}{r})^{-\epsilon-a}}dt^2+\frac{(l-y)^{\bar a}}{h^{a+b}(1-\frac{2M}{r})^{-\epsilon+a+b}}dr^2+\rho^2 d\Omega^2+R_c^2 dy^2,
\end{equation}
where the constant parameter $\epsilon\equiv\frac{k}{3M}$. I use the following definition for the proper radial coordinate: $\rho=r(l-y)^{\bar c}/h^{\frac{a+b-1}{2}}(1-\frac{2M}{r})^{\frac{-\epsilon-1+a+b}{2}}$, and for the "compactification radius": $R_c=h^{\frac{b}{2}}(l-y)^{\frac{\bar b}{2}}/(1-\frac{2M}{r})^{\frac{-\epsilon-b}{2}}$. The following constant redefinitions have been made: $\bar m\equiv ((2m/3-1)a+4mb/3-2m/3)/(a+2b-1)$, $\bar a\equiv ((2m/3+1)a+(4m/3+1)b-1)/(a+2b-1)$, $\bar b\equiv (2ma/3+(4m/3-1)b-2m/3)(a+2b-1)$, and $\bar c\equiv(2m/3+1)a+(4m/3+1)b-4m/3)/2a+4b-2$. For $\epsilon=0$ and on a given slice of fixed $y$-position in the extra space one recovers the soliton solution of Refs.\cite{wesson,gross-perry}. Since the proper radial coordinate should be real and positive definite it is apparent that, $r\geq 2M$. Therefore, the range allowed for the Schwarzschild radial coordinate $r\in [+\infty,2M]$. Besides, for $\epsilon<-1+a+b$, the proper radial coordinate $\rho$ is a minimum at $r^*=(-\epsilon+1+a+b)M$. From expression (\ref{sol1}) it is clear that there are not event horizons if, besides, $\epsilon<-a$. If the above constrains are fulfilled, then the curvature scalar,

\begin{equation}
\label{ricci1}
\hat R=-\frac{4}{3}\{\frac{k^2 h^{a+b}}{r^4}\frac{(1-\frac{2M}{r})^{-\epsilon-2+a+b}}{(l-y)^{\bar a}}+\frac{m^2}{h^b}\frac{(1-\frac{2M}{r})^{-\epsilon-b}}{(l-y)^{\bar b+2}}\},
\end{equation}
is bounded (and well behaved) for the range allowed for $r$. From this the wormhole structure is apparent: A throat of radius $\rho^*=(-\epsilon+1+a+b)^2 M (l-y)^{\bar c}/(-\epsilon-1+a+b)h^\frac{a+b-1}{2}$ joints two asymmetric asymptotic regions. The one at $r\rightarrow 2M$ is Ricci flat, meanwhile the one at $r\rightarrow\infty$ is not Ricci flat ($\hat R=-\frac{4}{3}\frac{m^2}{h^b}(l-y)^{-\bar b-2}$). When the above constrains are not fulfilled, i. e., when $\epsilon\geq -1+a+b$ one obtains a naked singularity at $r=2M$. There $\rho=0$ and, even if $\epsilon\geq -a$, the surface area of the "even horizon" (infinite redshift surface) is zero. Another important feature of this solution I want to note is the fact that, if $\epsilon<-b$, the "compactification radius" $R_c$ blows up to infinity as one approaches the spatial infinity at $r=2M$. In the case if one considers closed topology of the extra dimension (following the analysis in Ref.\cite{agnese}), this event opens the possibility that the unobservable compactified extra dimension(s) become visible and even dominate the physics operating near of the wormhole throat and behind it.

For $a=1$, $b=0$, the situation is very similar. In this case one has for the line-element the following expression,

\begin{equation}
\label{sol2}
ds_5^2=-\frac{e^{(\frac{2}{3}h+l)y}}{(1-\frac{2M}{r})^{-\epsilon-1}}dt^2+\frac{e^{(\frac{2}{3}h-l)y}}{(1-\frac{2M}{r})^{-\epsilon +1}}dr^2+\rho^2d\Omega^2+R_c^2 dy^2,
\end{equation}
where the following proper radial coordinate and "compactification" radius have been introduced: $\rho=re^{\frac{1}{3}hy}/(1-\frac{2M}{r})^{-\frac{\epsilon}{2}}$ and $R_c=e^{\frac{1}{3}hy}/(1-\frac{2M}{r})^{-\frac{\epsilon}{2}}$. The standard Schwarzschild metric (with an extra flat dimension) is recovered on a given slice of fixed $y$-position in the extra space if one sets in (\ref{sol2}) $\epsilon=0$. The range of the Schwarzschild $r$-coordinate is constrained to $r\in[+\infty,2M]$ by the fact that the proper radial coordinate $\rho$ shoul be real and positive. If the constrain $\epsilon<-1$ is fulfilled, then there are no horizons (see the time-time component of the metric in (\ref{sol2})) and, besides, the curvature scalar,

\begin{equation}
\label{ricci2}
\hat R=-\frac{4}{3} e^{-\frac{2}{3}h}\{\frac{k^2}{r^4}e^{ly}(1-\frac{2M}{r})^{-\epsilon-1}+h^2(1-\frac{2M}{r})^{-\epsilon}\},
\end{equation}
is bounded (and well-bahaved) in the range allowed for $r$. In this case the proper radial  coordinate is a minimum at $r^*=(-\epsilon+2)M$. One is faced with a wormhole with a throat of radius $\rho^*=(-\epsilon)^{\frac{\epsilon}{2}}(-\epsilon+2)Me^{\frac{h}{3}y}$, joining two asymmetric asymptotic regions: one Ricci flat at $r=2M$ and the other at $r=\infty$ that is not Ricci flat. For $\epsilon\geq -1$ one obtains clearly a naked singularity.

For testing the above geometry one can use a test particle (perhaps, uncharged and spinless). One can study the 5-dimensional geodesic motion of this particle that, given in a Weyl-integrable manifold, it is driven by the following equations: $\frac{d^2x^A}{ds_5^2}+\{^{\;\;A}_{MN}\}\frac{dx^M}{ds_5}\frac{dx^N}{ds_5}=\frac{1}{3}\frac{\phi_{,N}}{\phi}(g^{AN}-2\frac{dx^A}{ds_5}\frac{dx^N}{ds_5})$, and them one can check the ausence or presence of singularities in the Weyl- integrable manifold. In the case we are studying here the result is that, depending of the range chosen for the parameters of the solutions, one is able to describe two different situations: 1) The geodesics are complete; these go from one spatial infinity, approach the wormhole throat and then diverge into the second spatial infinity. 2) The geodesics end up in a naked singularity: both the proper time and the coordinate time taken for the test particle to fall from a given spatial point into the singular surface $r=2M$ are finite.

\section{Physical Discussion} 

The former results show that, if the present model is correct, there are two possible outcomes of the gravitational collapse: A wormhole and a naked singularity. However, when one deals with a compact extra dimension, then $R_c$ in Eqs. (\ref{sol1}) and (\ref{sol2}), has literally the meaning of a compactification radius. In consequence, the following pictures emerge: If $\epsilon>-b$ or $\epsilon>0$ depending on the case ($a\neq 1$, $b\neq 0$ or $a=1$, $b=0$), then, the compactification radius decreases to zero when one approaches the naked singularity. This seems to be in contradiction with the standard model picture, according to which, the extra dimensions decompactify (and, eventually, become very large) with increasing of the energy scales (for instance if one travels back in cosmological time into the early universe). Unlike this, in the contrary cases: $\epsilon<-b$ or $\epsilon<0$, the compactification radius $R_c$ increases while approaching the hypersurface $r=2M$. It seems likely that, in consequence, $\epsilon$ should be in the above range. If the following constrain holds; $-1+a+b\leq\epsilon<-b$ ($-1\leq\epsilon<0$ for the Schwarzschild case), then a naked singularity occurs. However this range for possible values of $\epsilon$ is much more restricted than the range in which wormhole configurations are favoured: $\epsilon<-1+a+b$ or $\epsilon<-1$ for the Scwarzschild case. There is yet the Penrose's Cosmic Censorship Conjecture favouring this last (wormhole) type of configuration.

In this context I want to discuss here on the null energy condition (NEC) that is always violated in an arbitrary wormhole throat\cite{barcelo}. Violation of NEC means that, even a null geodesic observer would see negative energy densities on passing the throat. However in a higher (five)-dimensional description, NEC violation is much less problematic. In effect, consider a 5-dimensional null-vector $k^A$ ($k_N k^N=0$) with components: $k^A=(\sqrt{-g^{00}},0,0,0,\sqrt{g^{55}})$ and consider, besides, that the NEC is violated: $\hat R_{NM}k^N k^M<0\Rightarrow\hat T_{NM}k^N k^M<0$. This last inequality can be rewritten (for a diagonal stress-energy tensor) in the following form: $\hat T_{00}(k^0)^2<-\hat T_{55}(k^5)^2$, or $g^{00}\hat T_{00}>g^{55}\hat T_{55}$. The LHS of this inequality can be interpreted as the energy density as measured by a four-dimensional "comoving" observer with time-like vector $v^a=(\sqrt{-g^{00}},0,0,0)$. In consequence, it is apparent that, if the RHS of this inequality is positive (for a spacelike extra dimension this implies a negative $\hat T_{55}$ component of the stress-energy tensor), then the energy density as measured by a 4-dimensional "comoving" observer could be positive and, in correspondence, no exotic 4-dimensional matter degrees of freedom are necessary at the wormhole throat. Even if the RHS of the above inequality is non-negative, some possibilities remain. Among them, the most attractive being Wesson's induced matter approach. Within the context of this last proposal, the observable 4-dimensional matter degrees of freedom are induced from "pure" geometry in the higher dimension. Therefore, the constraints on higher-dimensional (geometrical) magnitudes are not really constraints on the lower-dimensional matter degrees of freedom these induce. In particular, to the $(0,0)$-component of the 5-dimensional stress-energy tensor (which in the present model is given by: $\hat T_{AB}=-\frac{4}{3}\phi^{-2}[\phi_{,A}\phi_{,B}-\frac{1}{2}g_{AB}(\hat\nabla\phi)^2]+\phi^{-1}\hat\nabla_A\hat\nabla_B\phi$), it should be added an additional $T_{00}^{ind}$ term (in fact the energy density induced from the higher dimension), that should be positive definite irrespective of the sign of $\hat T_{00}$ thus warranting non-exotic matter content.

In regard to the geometrical structure of the manifold, the scalar (Weyl) function is given by $\phi=(1-2M/r)^{\frac{3}{2}\epsilon}(l-y)^m$ for $a\neq 1$, $b\neq 0$, and $\phi=(1-2M/r)^{\frac{3}{2}\epsilon}e^{hy}$ when $a=1$ and $b=0$. Therefore, in both cases the $r$-gradient of $\phi$ varies like: $\partial_r\phi\approx 3\epsilon M(1+(1-\frac{3}{2}\epsilon)2M/r)/r^2$. The result is that for $r>>2M$, $\partial_r\phi\sim 0$, i. e., far enough from the naked singularity or the wormhole throat (depending on the range of $\epsilon$ one chooses), $\phi$ is nearly constant and, in correspondence, the geometrical structure of the manifold is nearly Riemannian. In this case the present theory (action(\ref{bdaction})) approaches Einstein's theory given by the standard (five-dimensional) Einstein-Hilbert action.

\section{Conclusions}

The possibility that the destiny of the gravitational collapse could be a wormhole is very attractive since, then, the undesirable spacetime singularity is removed from the description of real physical processes involved. However, the plausibility of experimental testing is unclear since, the present approach could yield (perhaps appreciable) departures from standar Riemannian formulation, only in the neighbourhood of collapsed spacetime regions, so the solar system is not a good candidate for such testing. Nevertheless, this does not demerit the advantages of this model: It is simple (pure geometry in five dimensions, i. e., 5-dimensional vacuum) and classical (the action principle used does not consider quantum processes), but does not miss a point of the basic ideology of the unification scheme. Perhaps, assuming a Weylian structure of the underlying geometry, means that some quantum behaviour could be described since, as stated in the bibliography\cite{novello,dzhunushaliev}, Weyl structures mimic (classically) quantum behaviour. Besides, the possibility that there are macroscopic regions in our universe where scale invariance of the physical laws is preserved, seems very nice. 

The simplicity of the model is, at the same time, its main disadvantage: One knows that with increasing of energy scales, dimensionalities higher than five are expected to happen (10 for the String Theory and 11 for M-theory) and, besides, a fundamental theory in the spirit of String Theory or alike is to take place. Nevertheless, the present model is distinguished by its manageability and may account for a qualitative description of what one expects to occur during the last stages of the collapse.

In this direction, it will be interesting to have a brane model for a kind of description accounted for in the present paper. What one expects is that, far from the (collapsed) source, test (SM) particle's geodesics are trapped to a 3-brane (a 4-dimensional hypersurface embedded in the 5-dimensional "bulk"). But, as the SM particle's geodesic approaches the source, the particle's energy increases and, behind a given energy scale, the particle can adquire momentum in the extra direction and can scape into the bulk, thus leaving behind what a 4-dimensional observer living on the brane acknowledges as a singularity (an end point of a 4-dimensional geodesic).

Finally I want to stress that the kind of solutions presented here belongs to a wider class advanced in Ref.\cite{agnese}.


I acknowledge the MES of Cuba by financial support of this research.



\begin{thebibliography}{99}


\bibitem{hawking} S. W. Hawking and G. F. R. Ellis, {\it The Large Scale Structure of Space-Time} (Cambridge University Press, Cambridge, England, 1973).

\bibitem{singularity} S. K. Rama, Phys. Rev. Lett. {\bf 78}, 1620 (1997); Phys. Lett. B{\bf 408}, 91 (1997); N. Kaloper and K. A. Olive, Phys. Rev. D{\bf 57}, 811 (1998); C. Park and S-J. Sin, Phys. Rev. D{\bf 57}, 4620 (1998);  J. E. Lidsey, D. Wands and E. J. Copeland, Phys. Rept. {\bf 337}, 343 (2000); I. Quiros, Phys. Rev. D{\bf 61}, 124026 (2000).

\bibitem{novello} M. Novello, L. A. R. Oliveira and J. M. Salim, Int. J.
Mod. Phys. D, Vol. 1, Nos 3 \& 4 (1993)641.

\bibitem{arkani-hamed} N. Arkani-Hamed, S. Dimopoulos and G. Dvali, Phys. Lett. B{\bf 429}, 263 (1998); I. Antoniadis, N. Arkani-Hamed, S. Dimopoulos and G. Dvali, Phys. Lett. B{\bf 436}, 257 (1998); N. Arkani-Hamed, S. Dimopoulos and G. Dvali, Phys.Rev. D{\bf 59}, 086004 (1999); L. Randall and R. Sundrum, Nucl. Phys. B{\bf 557}, 79 (1999); L. Randall and R. Sundrum, Phys. Rev. Lett. {\bf 83}, 3370 (1999); L. Randall and R. Sundrum, Phys. Rev. Lett. {\bf 83}, 4690 (1999); J. Lykken and L. Randall, JHEP {\bf 0006}, 014(2000).

\bibitem{horava-witten} P. Horava and E. Witten, Nucl. Phys. B{\bf 460}, 506 (1996); E. Witten, Nucl. Phys. B{\bf 471}, 135 (1996); P. Horava and E. Witten, Nucl. Phys. B{\bf 475}, 94 (1996).

\bibitem{duff} M. J. Duff, Int. J. Mod. Phys. A{\bf 11}, 5623 (1996).

\bibitem{witten} M. B. Green, J. H. Schwarz and E. {\it Superstring Theory} (Cambridge University Press, 1987). 

\bibitem{kk} T. Kaluza, Sitz. Preuss. Akad. Wiss. Phys. Math. K{\bf 1}, 966 (1921); O. Klein, Zeits. Phys. {\bf 37}, 895 (1926); M. J. Duff, hep-th/9410046; J. M. Overduin, P. S. Wesson, Phys. Rept. {\bf 283}, 303 (1997).

\bibitem{weyl} H. Weyl, {\it Space, Time and Matter} (Dover, NY, 1952); R. Adler, M. Bazin and M. Schiffer, {\it Introduction to general relativity}, 2d ed. (McGraw-Hill, New York, 1975).

\bibitem{dzhunushaliev}  V. Dzhunushaliev and H.-J. Schmidt, Phys. Lett. A{\bf 267}, 1(2000).

\bibitem{sato} A. Tomimatsu in H. Sato and T. Nakamura (Eds.), {\it Gravitational Collapse and Relativity}, p. 417 (World Scientific, 1986)

\bibitem{wesson} P. S. Wesson, J. Ponce de Leon, P. Lim and H. Liu, Int. J. Mod. Phys. D{\bf 2}, 163 (1993); P. S. Wesson, {\it Space-Time-Matter: Modern Kaluza-Klein Theory}, (World Scientific Publishing Co., 1999).

\bibitem{gross-perry} D. J. Gross and M. J. Perry, Nucl. Phys. B{\bf 226}, 29 (1983); D. Davidson and D. Owen, Phys. Lett. B{\bf 155}, 247 (1985).

\bibitem{agnese} A. G. Agnese and M. La Camera, Phys. Rev. D{\bf 58}, 087504 (1998).

\bibitem{barcelo} C. Barcelo and M. Visser, Phys. Lett. B{\bf 466}, 127 (1999).
 

\end{thebibliography}
\end{document}